# permeability distribution of gas drainage of borehole with the different moisture content caused polar permeability effect

PAN Hongyu[1], ZHANG Yao[1], CAO Yan[1], CHU Yuhang[1], YANG Shihua[1], ZHANG Lei[1],
(1.College of Safety Science and Engineering, Xi'an University of Science and Technology, Xi'an Shan xi 710054, China;)

**Abstract:** In order to study the penetration characteristics in areas with different water content and different stress distributions in the radial direction of the hole after hydraulicization measures, an improved LFTD1812 triaxial permeability meter was used to conduct a test to measure the polar permeability characteristics of coal with different water content combinations were measured by permeability instrument, and the porosity, permeability, pressure gradient and seepage velocity of different samples were analyzed. The relationship between sample porosity, permeability, pressure gradient and seepage velocity was discussed, the influence of moisture content on permeability was discussed, and the directionality and the directivity and polarization effect of permeability were found.. Result shows that The relationship between permeability and porosity shows two trends of exponential type and logarithmic type, and the porosity-permeability($\varphi$-$k$) plane is divided into three influence regions: super index (I), index (II) and logarithm (III). The permeability of the sample shows an exponential decay trend of slow and rapid stages with the increase of the absolute value of the pressure gradient, and the higher the water content is, the lower the initial permeability is, and the permeability-pressure gradient characteristic curve moves downwards as a whole. the slow decline zone moves to the right, and the absolute value of the pressure gradient required to enter the slowly decreasing area decreases, and the seepage velocity decreases with the increase of water content at all levels of osmotic pressure, and the decreasing trend gradually weakens. The directivity of permeability weakens with the increase of equivalent water content, osmotic pressure and axial pressure, and water content is the main influence factor of directionality, and the polarization effect shows three modes: weak appearance, transition appearance and strong appearance under the influence of water cut. There are two states of incomplete polarization and complete polarization, which can weaken the direction of permeability. The study can provide reference for the study of the influence of moisture on coal permeability, and provide theoretical support for the application of hydraulic measures in coal mine gas prevention and control.At present, coal still occupies the main position of China 's energy structure. In order to promote the completion of China 's dual carbon goal in 2050, it is necessary to increase the production of unconventional natural gas. Hydraulic measures are used to increase the production of unconventional natural gas, which mainly changes the permeability of coal seam after water injection, but the influence of water after water injection on unconventional natural gas extraction is not clear.

## 1 Introduction

There are a large number of high gas mines in China, and coal seam pre-drainage measures are widely adopted to prevent the occurrence of gas accidents [1]. With the increase of mining depth, the ground stress increases and the permeability of coal seam decreases, so it is necessary to further strengthen the extraction. To this end, since the 1970 s, a variety of experiments have been carried out to strengthen drainage, and a variety of hydraulic permeability measures have been developed. These hydraulic measures can be roughly divided into two categories according to the anti-reflection mechanism :pressurization class such as hydraulic fracturing [2], pressure relief class such as hydraulic slotting [3], hydraulic punching [4], etc. They have achieved good application results. However, after the implementation of hydraulic measures, the permeability of coal seam will increase significantly in the short term, and the long-term impact is more complicated.

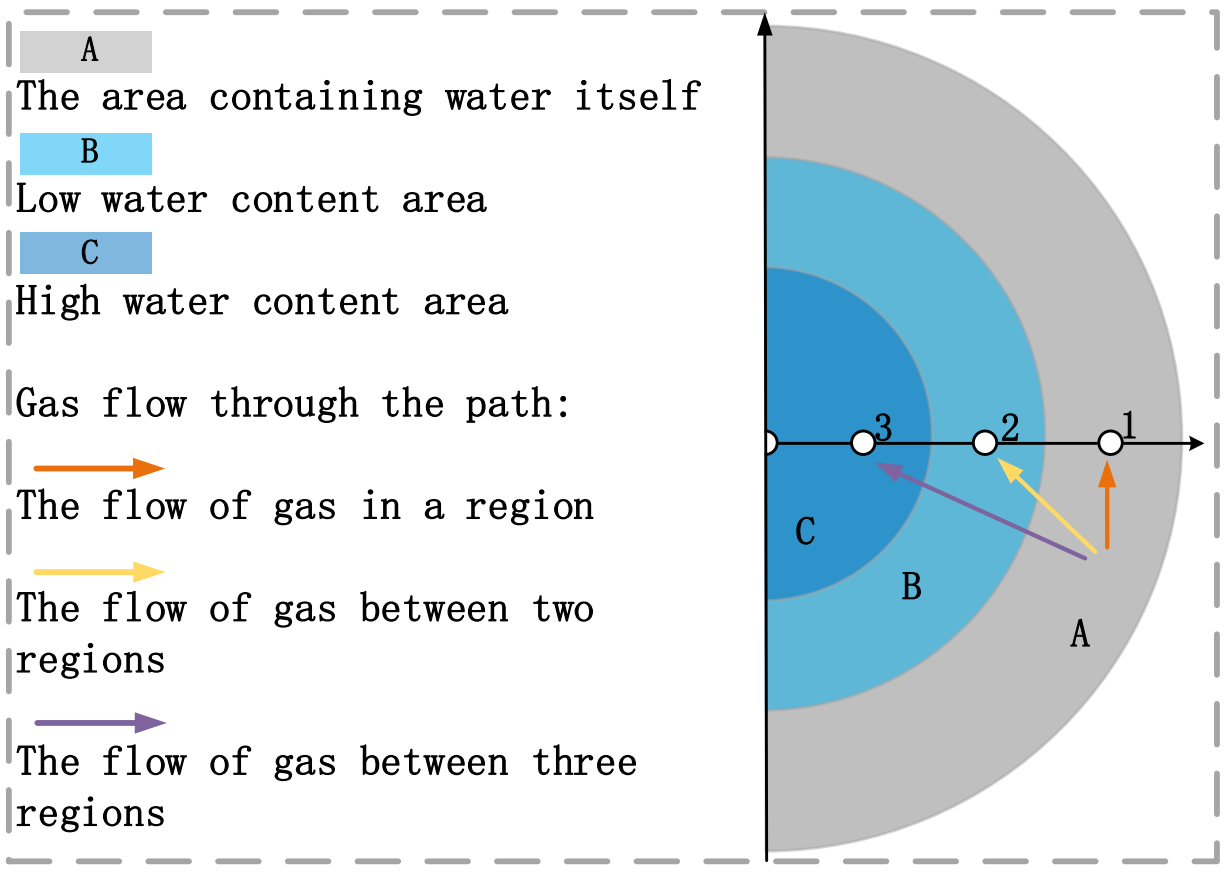

Fig.1 Schematic diagram of percolation path around hole

The reason is that under the action of water flow in the coal body around the hole, three different water content areas will be formed in the cracks of the coal body around the hole, and the flow of gas will be significantly affected (Fig.1). The flow of gas in the three regions can be divided into three categories. These three categories are Type I : single regional flow with constant water content ; type II : moisture content from low to high or from high to low double area flow ; type III : water content from low to medium to high or from high to medium to low three regional flow. Type I flow corresponds to the gas flow near the borehole in the high water content area, the borehole in the low water content area and the borehole in the water content area in the extraction project ; type II flow corresponds to the gas from the adjacent high water content area, low water content area or itself containing water area across a region into the region drilling ; type III flow corresponds to the drilling into the region from the far high water content area or the water-containing area itself across two regions. The flow in these areas involves the study of water content and permeability characteristics of coal.

In terms of the permeability characteristics of coal, In order to study the effect of effective stress on the permeability of high-rank coal, Zhang et al.[5] In order to quantitatively describe the migration and morphology of oil and gas in mining fractures, they used a self-developed online low-field nuclear magnetic resonance triaxial seepage system to conduct a triaxial seepage experiment considering mining redistribution stress. Based on T2 spectrum, MRI and stress-strain response, the evolution of stress-crack-seepage field is monitored and its coupling mechanism is analyzed. Zhu et al.[ 6 ] conducted permeability tests with different gas pressures and found that as the gas pressure increases, the permeability first decreases and then increases. Peng et al.[7] In order to study the impact mechanism of gas seepage on coal and gas outburst disasters. Wang et al.[8] In order to study the seepage and deformation characteristics of coal at high temperatures, CT scanning was used, combined with ANSYS software, to obtain the seepage and deformation characteristics of coal at high temperatures. Zhang et al.[9], In order to study the effect of loading rate on gas seepage and temperature, seepage tests at different loading rates were carried out, and it was concluded that stress directly affects the seepage speed and temperature of gas-containing coal. Gas pressure and gas content are the most sensitive to mining stress; temperature is the most sensitive to mining stress. Not sensitive. Meng et al. [10] In order to study the seepage rules of raw coal under different gas pressures, a self-developed gravity constant load seepage experimental system was used to carry out different gas pressure permeability tests, and the seepage rules of raw coal under the action of He, $N_2$ and $CO_2$ were obtained. Zhu et al. [11] In order to study the creep seepage law of coal and rock, the cscg-160 gravity hydraulic constant load creep seepage test system was used to conduct an axial load step loading creep seepage test on raw coal and rock samples under certain confining pressure conditions, and the creep seepage of coal rock was obtained. Creep deformation-seepage law in the process. Li et al. [12] In order to study the fluid seepage mechanism of multi-coal interbedded coal systems, numerical simulations and physical simulation experiments were combined to establish a sedimentation-structure controlled permeability prediction model. Pang et al. [13] In order to study the accurate measurement of permeability parameters, a self-designed triaxial permeability device for broken particle media was used to conduct experiments to determine the permeability rules of gas in the broken coal around the borehole, and the deformation and destruction process of the broken coal particles was divided into three stages.Liu et al.[14, 15] In order to characterize the relationship between fracture pore parameters and permeability, et al. established a coalbed methane seepage model based on the fractal structure characteristics of fractures and pores. This model can better characterize the gas drainage effect; thick capillary tube bundles were used to characterize the coal body structure. Based on the physical model, a permeability model including the fractal dimension of pore curvature and specific surface area was established.

The above studies show that different water content has a significant effect on gas seepage. However, after the hydraulicization measures, the flow patterns in three different water-bearing areas around the hole have not been systematically studied. And the seepage of three types of gas in three regions will have different forms. This involves the distribution of water content and the permeability test of multi-component coal under the influence of different water content. Therefore, it is necessary to measure the permeability characteristics of different water-bearing combination samples and consider their directivity, so as to describe the seepage characteristics of the water-bearing area around the hole.

Therefore, this paper carried out the determination test of polar permeability characteristics of coal samples with different water content combinations. The relationship between porosity, permeability, pressure gradient and seepage velocity of different samples is analyzed, and the relative permeability of water content is obtained. In order to provide reference for clarifying the influence of water content on the permeability of three types of three zones of coal body around the hole, and finally provide theoretical support for the application of hydraulic measures in coal mine gas prevention and control.

## 2 Theory

(1) Seepage velocity and pressure gradient

In the test, the seepage velocity is calculated according to the following equation [16]:

$$v = \frac{4Q}{\pi d^2} \qquad (1)$$

Where: $v$ is the seepage velocity; $Q$ is the flow rate through the sample; $d$ is the sample diameter.

The pressure gradient at both ends of the sample [17]

can be expressed as:

$$G_P = \frac{P_2 - P_1}{h_0\text{-}\Delta h} \tag{2}$$

Where: $G_P$ is the pressure gradient; $P_2$ is the pressure at the outlet end of the osmometer; $P_1$ is the pressure at the intake end of the osmometer; $h_0$ is the sample height, $\Delta h$ is the displacement change.

(2) Permeability

According to the isothermal gas seepage process, the gas seepage in the coal sample conforms to Darcy's law, then the permeability is calculated as [18]:

$$k = \frac{Q_0 \mu h_0}{A P_1} \tag{3}$$

Where: $k$ is gas permeability; $Q_0$ is the gas flow under atmospheric pressure; $\mu$ is the gas viscosity coefficient. $A$ is the cross-sectional area of the sample.

(3) Porosity

In the test, according to the relationship between axial displacement and porosity, the immediate porosity at every moment in the permeation process is [19]:

$$\varphi = 1 - \frac{m}{\rho_c A\left(h_0 - \Delta h\right)} \tag{4}$$

Where: $\varphi$ is the porosity; $\Delta h$ is the axial deformation height of the broken sample; $A$ is the cross-sectional area of the specimen in the additional device for the penetration test of the broken specimen; $m$ is the mass of the sample loaded in the test; $\rho_c$ is the mass density of coal particles. $h_0$ is the initial height of the sample.

(4) Equivalent moisture content

In order to facilitate the representation of the water content of the composite sample, the equivalent water content $\bar{\omega}$ is introduced, which represents the weighted average water content of each part, calculated as follows:

$$\bar{\omega} = \frac{\omega_i \cdot h_i}{\sum h_i} \quad (i = 1, 2, 3) \tag{5}$$

Where: $\omega_i$ is the moisture content of the part i of the sample; $h_i$ is the height proportion of part i of the sample; $i$=1, 2 and 3 represent the upper, middle and lower parts of the sample respectively.

(5) Effective stress

According to Terzaghi's principle, the equation for calculating effective stress $\sigma_e$ [20] is as follows:

$$\sigma_e = \frac{\sigma_z + 2\sigma_r}{3} - \frac{P_1 + P_2}{2} \tag{6}$$

Where: $\sigma_z$ and $\sigma_r$ are axial and confining pressures of coal samples respectively. $P_1$ and $P_2$ are gas pressures at both ends of coal sample inlet and outlet respectively.

## 3 Experimental design

### *3.1 Experimental equipment*

The improved LFTD1812 automatic triaxial permeameter was used in the test. Its main structure includes : confining pressure system, axial pressure system, osmotic pressure system, computer control system and MF5706-N-10 flowmeter. The confining pressure system is composed of an independent pressure pump and an additional pipeline. The axial compression system is responsible for providing the vertical pressure of the sample. The osmotic pressure system is composed of a high-pressure gas cylinder, an adjustable precision valve and an additional pipeline. The pipeline is connected to the upper and lower ends of the sample to provide the osmotic pressure difference at both ends of the sample. The computer control system is responsible for collecting axial and confining pressure data. The osmotic pressure is recorded by the precision valve of the high pressure cylinder, and the gas flow through the sample is recorded by the MF5706-N-10 flowmeter. The connection of the whole triaxial seepage test device is shown in Fig.2.

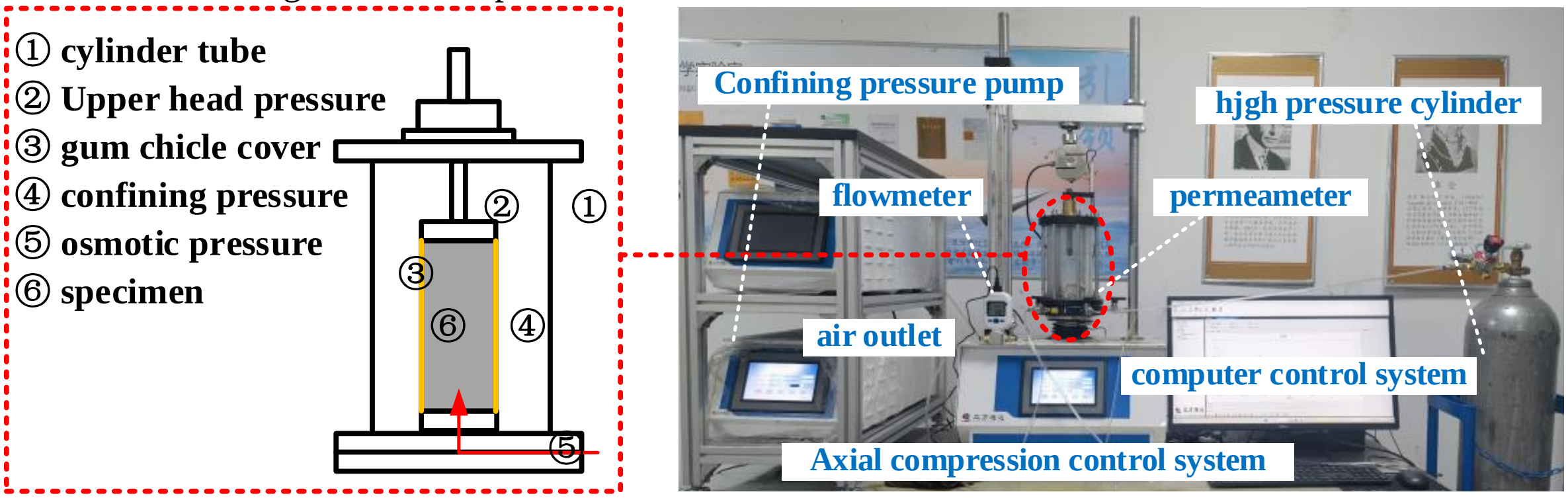

Fig.2 Experimental diagram of triaxial seepage devic

### *3.2 Sample design*

Because the hydraulic measures change the natural water content state around the extraction hole, the water content distribution around the hole is more complicated due to the influence of coal cracks. In order to facilitate the study, the water content distribution around the hole is simplified into three-zone distribution, as shown in Fig.1.

In Fig.1, the hole is divided into three regions. These three areas are : high water content area, low water content area, and area containing water itself. Their corresponding water content is assumed to be $\omega_1$, $(\omega_1 + \omega_2)/2$, $\omega_2$. According to the distribution of water content around the hole, the path of gas entering the extraction borehole from the coal seam can be divided into three categories : Type I, flow in a single region, Type II, flow between two regions, Type III, flow through three regions. The flow sample in a single area simulates the gas migration under the condition of uniform water content around the hole. The flow sample between the two regions simulates the gas migration in the case of high and low water content around the hole. The samples flowing through three regions simulated the gas migration under the two conditions of gradual increase or decrease of water content around the hole. The two-zone flow test provides a theoretical analysis basis for the three-zone flow test.

According to the field situation, the highest water content in the coal seam can reach 30 %, and the lowest natural water content can reach 0.8 % [21]. Due to the limitation of sample preparation conditions, the highest moisture content in the sample is set to 25 % ; the minimum part is set to 10 %. Therefore, the samples of different moisture content components were 25 %, 17.5 % and 10 %, respectively.

### *3.3 Sample preparation*

The coal sample was taken from the 13200 mining face of Gengcun Coal Mine, after crushing, weighing, shaping, drying, combination and other processes. The raw coal was crushed by CHIGO-1000 g-N type coal crusher, and the pulverized coal with particle size of 0 ~ 0.2 mm was screened. In weighing, it is required to obtain a certain quality of pulverized coal and the corresponding quality of distilled water, and the quality of pulverized coal $m_f$ and distilled water $m_w$ are respectively recorded and fully mixed. In the molding, the mixed slurry is mixed and molded(The height of type I sample is 100 mm, the height of type II sample block is 50 mm, and the height of type III sample block is 33.3 mm.), and a cylinder with a diameter of 50 mm is made according to the required height. In the drying process, the finished block was placed in a constant temperature and humidity box for drying, and weighed every 10 min. When the moisture content reached the predetermined value, the drying was completed. At this time, the mass of the coal sample is recorded as $m_0$, and the final moisture content of the coal sample $\omega$ is :

$$\omega = \frac{m_w}{m_0} \tag{7}$$

In the combination, blocks or pulverized coal with different moisture contents were added in turn according to the sample design, and the mass of the combined coal sample was recorded as $m$. Finally, the parameters of briquette specimens with different water content were prepared as shown in table 1. In particular, it is assumed that the distribution of water content along the gas seepage direction from low to high is positive sequence permeability, and from high to low is reverse sequence permeability.

### *3.4 Experimental procedure*

In order to simulate the bearing environmental conditions of broken coal, the confining pressure was set to be 400 KPa, and the axial pressure was loaded from 300 N to 900 N to measure the change of permeability of briquette samples. The specific steps of the test are as follows :

① Return to zero : adjust the testing machine to the initial state, check whether the test equipment and instrument are working properly ;

② Install the sample : The prepared sample is placed on the testing machine, and the upper and lower parts are tightened with multiple double-strand rubber bands to prevent air leakage ; place gauze to prevent pulverized coal from blowing into the flowmeter ;

③ Adjust the indenter : adjust the position of the axial compression loading indenter and set the axial displacement to zero ;

④ Loading confining pressure : The confining pressure is loaded to the set value, and the axial pressure is adjusted to the set value.

⑤ Gas seepage : adjust the pressure valve, when the gas pressure at the inlet end is stable, open the inlet valve, pass the gas, and measure the flow rate ;

⑥ According to the test plan, repeat 3-5 steps until the end of this group of tests.

## 4 Results

Combined with the equation (1) - (3) in section 2, the results of seepage characteristic parameters such as seepage velocity, pressure gradient and permeability are shown in table 1.

Table 1 Parameters of three types of experimental groups and parameters of seepage characteristics($P_{axial}$=700N,$P_2$=0.025MPa)

| | | Sample name | $m_{cl}$/g | $m_w$/g | $m_0$/g | $\omega$/% | $G_P$/MPa | $v$/(cm/s) | $k$（$10^{-12}$）/(m$^2$) |
|---|---|---|---|---|---|---|---|---|---|
| Type I | 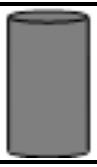 | H25G1 | 210 | 70 | 280 | 25 | -0.0265 | 0.8662 | 5.5514 |
| | | H175G1 | 210 | 44.55 | 254.55 | 17.5 | -0.0268 | 1.1125 | 7.0473 |
| | | H10G1 | 210 | 23.3 | 233.3 | 10.0 | -0.0273 | 2.1741 | 13.5443 |
| Type II | 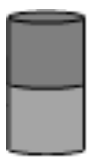 | H25175G11 | 105 | 35 | 140 | 25.0 | -0.0266 | 0.9597 | 6.1330 |
| | | | 105 | 22.27 | 127.27 | 17.5 | | | |
| | | H17510G11 | 105 | 22.27 | 127.27 | 17.5 | -0.0268 | 1.8599 | 11.7550 |
| | | | 105 | 11.67 | 116.67 | 10.0 | | | |

| | | | | | | | | |
|---|---|---|---|---|---|---|---|---|
| | | H17525G11 | 105 | 22.27 | 127.27 | 17.5 | -0.0263 | 1.3929 | 6.8081 |
| | | | 105 | 35 | 140 | 25.0 | | | |
| | | H10175G11 | 105 | 11.67 | 116.67 | 10 | -0.0263 | 1.7919 | 11.5872 |
| | | | 105 | 22.27 | 127.27 | 17.5 | | | |
| | | H25175G12 | 70 | 23.33 | 93,33 | 25 | -0.0264 | 1.0021 | 6.4453 |
| | | | 140 | 29.7 | 169.7 | 17.5 | | | |
| | 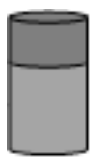 | H17510G12 | 70 | 14.85 | 84.85 | 17.5 | -0.0265 | 2.1401 | 13.7238 |
| | | | 140 | 15.56 | 155.56 | 10 | | | |
| | | H17525G12 | 70 | 14.85 | 84.85 | 17.5 | -0.0263 | 1.6730 | 6.6419 |
| | | | 140 | 46.67 | 186.67 | 25 | | | |
| | | H10175G12 | 70 | 7.78 | 77.78 | 10 | -0.0272 | 1.6221 | 10.1164 |
| | | | 140 | 29.7 | 169.7 | 17.5 | | | |
| | | H25175G21 | 140 | 46.67 | 186.67 | 25 | -0.0269 | 0.9172 | 5.7969 |
| | | | 70 | 14.85 | 84.85 | 17.5 | | | |
| | 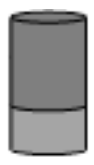 | H17510G21 | 140 | 29.7 | 169.7 | 17.5 | -0.0262 | 1.2739 | 8.2494 |
| | | | 70 | 7.78 | 77.78 | 10 | | | |
| | | H17525G21 | 140 | 29.7 | 169.7 | 17.5 | -0.0268 | 1.2824 | 8.1276 |
| | | | 70 | 23.33 | 93.33 | 25 | | | |
| | | H10175G21 | 140 | 15.56 | 155.56 | 10 | -0.0263 | 2.2217 | 14.2953 |
| | | | 70 | 14.85 | 84.85 | 17.5 | | | |
| Type III | 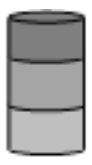 | H2517510G111 | 70 | 23.33 | 93.33 | 25 | -0.0264 | 1.2739 | 8.2088 |
| | | | 70 | 14.85 | 84.85 | 17.5 | | | |
| | | | 70 | 7.78 | 77.78 | 10 | | | |
| | | H1017525G111 | 70 | 7.78 | 77.78 | 10 | -0.0265 | 1.4013 | 8.9929 |
| | | | 70 | 14.85 | 84.85 | 17.5 | | | |
| | | | 70 | 23.33 | 93.33 | 25 | | | |

### *4.1 Porosity characteristics in the seepage process*

According to the equation (4), the change of axial pressure and porosity is drawn as Fig.3.

Through the observation of Fig.3 (a) and (c), it is found that the porosity decreases with the increase of axial pressure, and the law is consistent under different water content. The higher the water content, the lower the porosity, and the linear relationship between the increase of water content and the decrease of porosity is not obvious. From the perspective of the degree of change in porosity, the decrease in porosity of type III samples is weaker than that of the other two groups, showing a relatively gentle curve. This may be due to the fact that the initial structure of different blocks is closer than that of type I and type II samples after multiple compactions during the preparation of type III composite samples, so the porosity changes less under the same axial pressure.

By observing Fig.3 (b) and (c), it is found that with the increase of axial pressure, the porosity of the composite samples with different height ratios decreases with the increase of axial pressure. The larger the proportion of high water content in the composite sample, the lower the initial porosity of the sample. In general, the higher the equivalent water content of the sample, the lower the initial porosity.

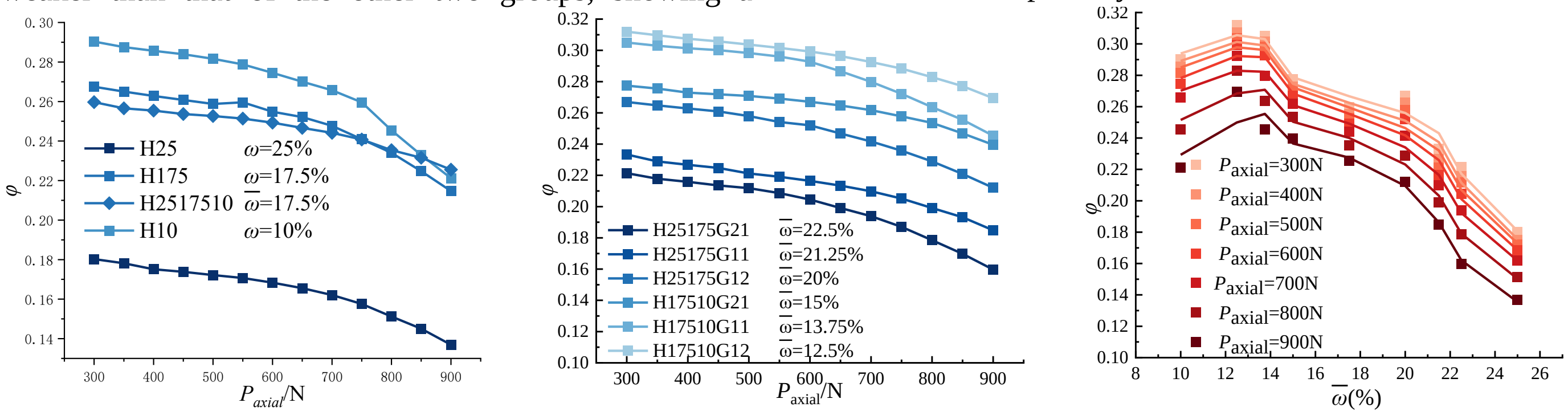

(a) Standard samples with different moisture content $P_{\text{axial}}$-$\varphi$

(b) Composite specimens of different height proportions $P_{\text{axial}}$-$\varphi$

(c) $\overline{\omega}$ -$\varphi$

Fig.3 Pore characteristic curves of different samples

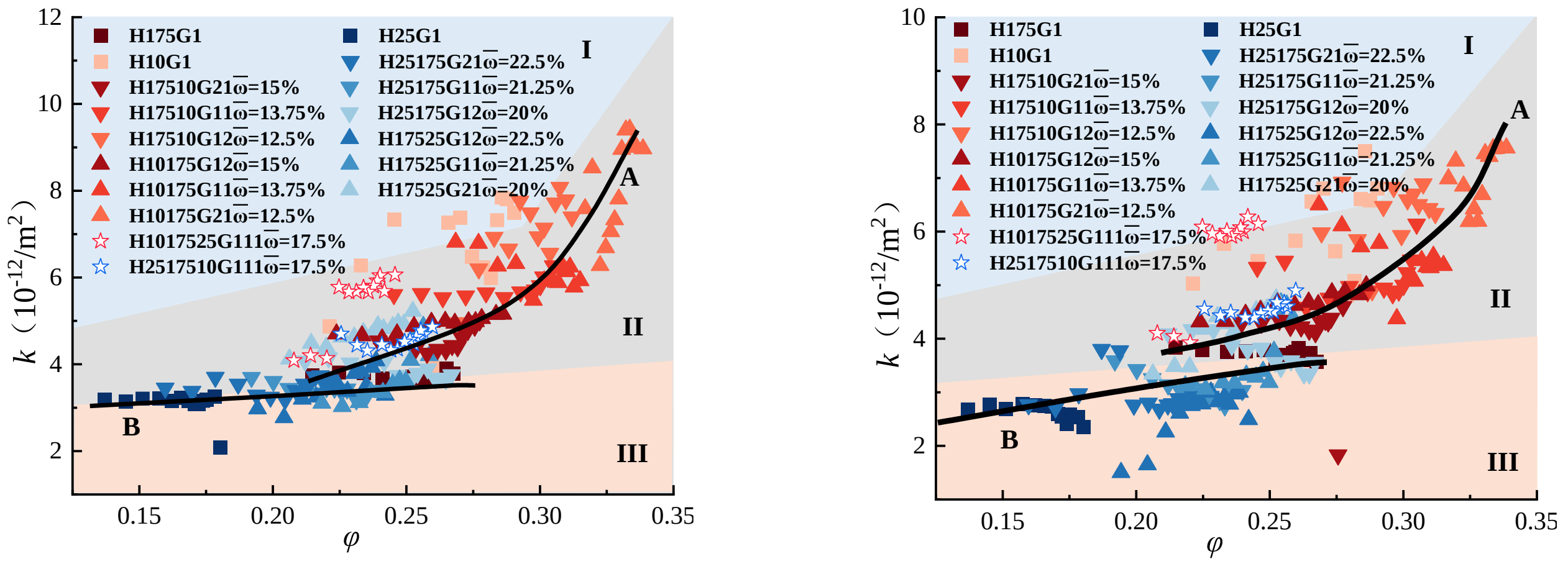

(a) $P_2$=0.01MPa (b) $P_2$=0.16MPa

Fig.4 Porosity - permeability curves of samples with different water content

Comparing Fig.3 (a) with (b), it is found that the porosity of all samples is consistent with the change of axial pressure, and the equivalent water content of the type II combination sample is similar to the influence of the water content of the type I sample on the initial porosity.

These results are basically consistent with the literature [22], and the equivalent water content of the composite sample is similar to the effect of the water content of the standard sample on the porosity, and the higher the water content is, the lower the initial porosity is. This is because with the increase of water content in the test, a large amount of free water is produced in the skeleton of the sample, which moves freely in the coal body and squeezes the seepage channel of the original gas.

Through the observation of Fig 4, it is found that the permeability generally shows two nonlinear increasing trends of exponential type (A) and logarithmic type (B) with the increase of porosity. These two growth curves divide the φ-k plane into three regions : I super exponential influence, II exponential influence, and III logarithmic influence. Among them, the permeability of the sample in the I super-index affected area shows a super-index change relationship with the increase of porosity ; the permeability of the samples in the area affected by the II index shows an exponential growth trend with the increase of porosity. III The permeability of the sample in the logarithmic influence area shows a logarithmic growth trend with the increase of porosity.

These areas are related to the abundance of structural water and free water. When the water content in the composite samples with different height ratios is mostly 25 %, the growth trend shows logarithmic growth. When the water content is mostly 10 %, its growth trend shows exponential growth. Specifically, when the water content or equivalent water content is less than 17.5 %, its growth trend shows a logarithmic growth ; when the water content or equivalent water content is greater than or equal to 17.5 %, the growth trend shows an exponential growth.

It is also found in the observation figure that some data of H10G1 group fall in the I-excess index influence area. This may be due to the fact that during the drying process of the sample, the internal structural water is lost with the drying part, and the pore structure in the sample increases, forming more seepage channels, so that the permeability of the sample is larger than that in the II-excess index influence area.

### *4.2 Influence of pressure gradient on permeability characteristics*

According to the change of flow rate with osmotic pressure measured in the experiment, the seepage velocity and pressure gradient are calculated by equation (1)-(3), and the relationship between seepage velocity and permeability and pressure gradient of three kinds of samples is drawn, as shown in Fig.5 and Fig.6 respectively.

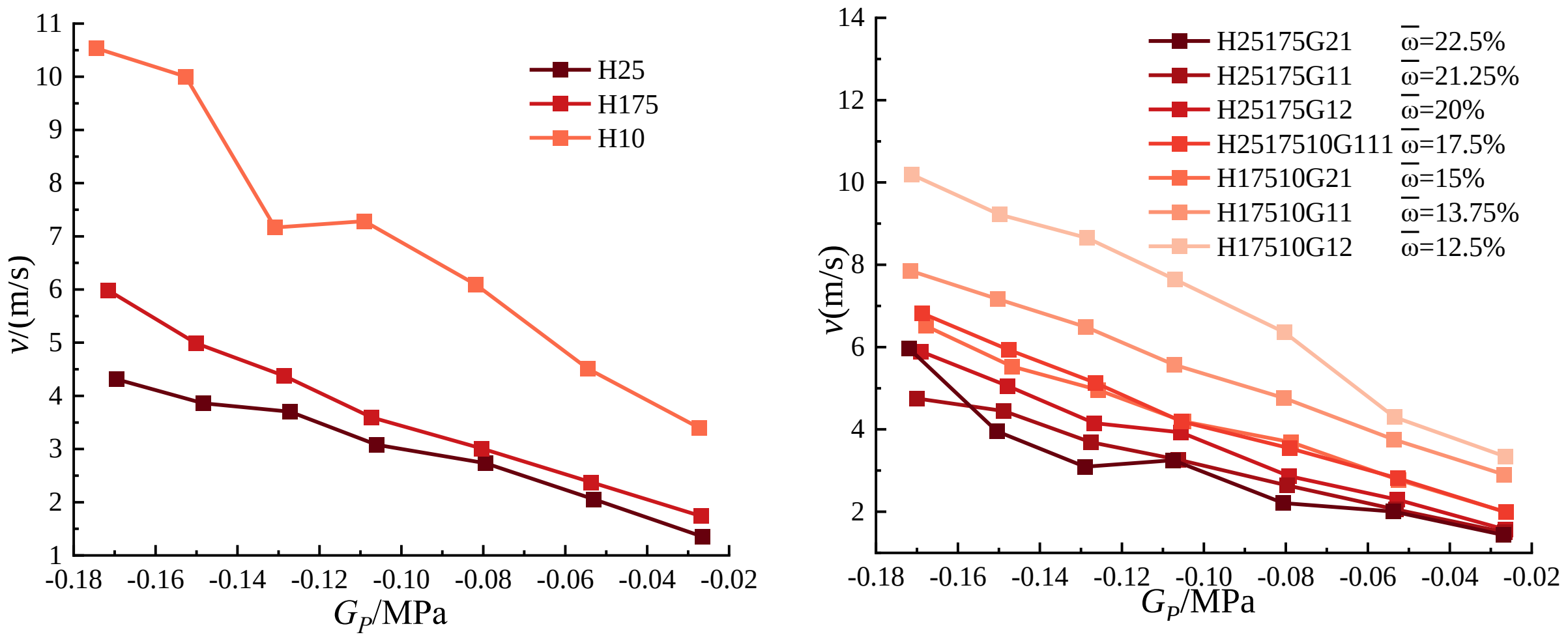

(a) $P_{axial}$=700N (b) $P_{axial}$=700N

Fig.5 Flow-pressure gradient curves of different samples

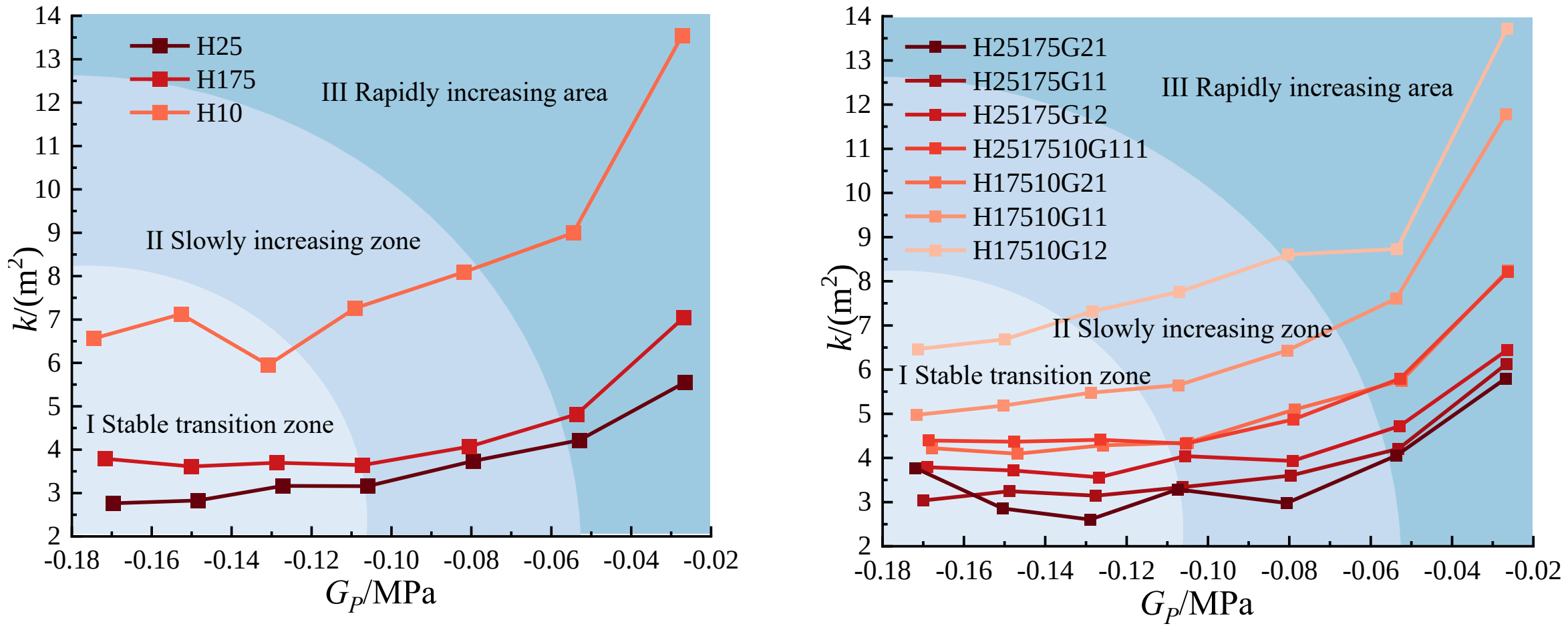

(a) $P_{axial}$=700N (b) $P_{axial}$=700N

Fig.6 Pressure gradient-permeability curves of different samples

(1) Influence of pressure gradient on seepage velocity

It is found from Fig.5 (a) that under the same osmotic pressure, the seepage velocity of standard samples with different water contents is negatively correlated with the pressure gradient, and the seepage velocity decreases with the increase of water content under all levels of osmotic pressure. This is consistent with the result in Reference [23] that the absolute value of the pressure gradient increases with the increase of the flow rate, but with the increase of the water content, the decreasing trend of the seepage velocity of the sample under the same pressure gradient is weakened. This means that the increase of water content can weaken the gas permeability process, reflecting the inhibitory effect of water on the seepage, which may be one of the reasons for the water lock effect in the gas permeability enhancement project.

It is found from Fig.5 (b) that the seepage velocity of the composite specimens with different height ratios is also negatively correlated with the pressure gradient. In the composite sample, the larger the proportion of high water content, the smaller the seepage velocity, and the inhibitory effect of water on the seepage velocity increases with the increase of the proportion of high water content. However, there is a case where the three-group sample with an equivalent water content of 17.5 % intersects with the two-group sample H17510G21 with an equivalent water content of 15 %. This may be due to the fact that in the three-group sample, as the pressure gradient reaches-0.08 MPa / m, the increase of osmotic pressure causes a penetrating channel inside the sample, and the permeable medium is connected to the part of the sample with a water content of 25 %. At this time, with the increase of osmotic pressure, part of the water in the sample will be taken out, forming a channel through the whole sample, resulting in the increase of seepage velocity, showing the polarization effect of seepage.

(2) Influence of pressure gradient on permeability

It is found from Fig.6 (a) that the permeability of samples with different water contents decreases with the increase of the absolute value of pressure gradient, showing an exponential decay trend, and showing three stages of smooth transition, slow increase and rapid increase. In the test, the permeability k does not increase proportionally with the increase of osmotic pressure, and the higher the water content is, the lower the initial permeability is. The permeability-pressure gradient characteristic curve moves downward as a whole, and all stages move right. The absolute value of the pressure gradient in the area where the permeability of the sample enters a slowly decreasing area decreases. Reference [24] believed that with the increase of osmotic pressure, the permeability decreased first and then increased, with obvious Klinkenberg effect. However, in this experiment, the Klinkenberg effect only appeared at the end of the test of some samples. This may be due to the low loading value of osmotic pressure in this test, which did not reach the threshold of permeability increase.

It is found from Fig.6 (b) that the permeability of the combined samples with different height ratios also has an exponential attenuation trend that decreases with the increase of the absolute value of the pressure gradient, and the more the high water content is, the lower the initial permeability is. The permeability-pressure gradient characteristic curve moves down as a whole, and all stages move to the right, so that the absolute value of the pressure gradient that slowly reduces the permeability of the sample decreases. Literature [25] found that with the increase of water content of the sample, the permeability showed a decreasing trend under the same osmotic pressure. In this paper, it is found that the three-group sample with equivalent water content of 17.5 % and the two-group sample with equivalent water content of 15 % do not strictly follow this law. After the pressure gradient reaches − 0.08 MPa / m, the curves intersect. Under the same pressure gradient, the permeability of the three-group sample is higher than that of the two-group sample, which may also be caused by the above polarization effect.

### *4.3 Influence of (equivalent) moisture content on permeability of coal*

According to equation (3) and equation (5), the equivalent water content and permeability are calculated, and the relationship between permeability and equivalent water content under different axial pressures is drawn as shown in Fig.7. Among them, ω = 10 %, 17.5 %, and 25 % are class I standard samples, and the rest are class II double combination samples.

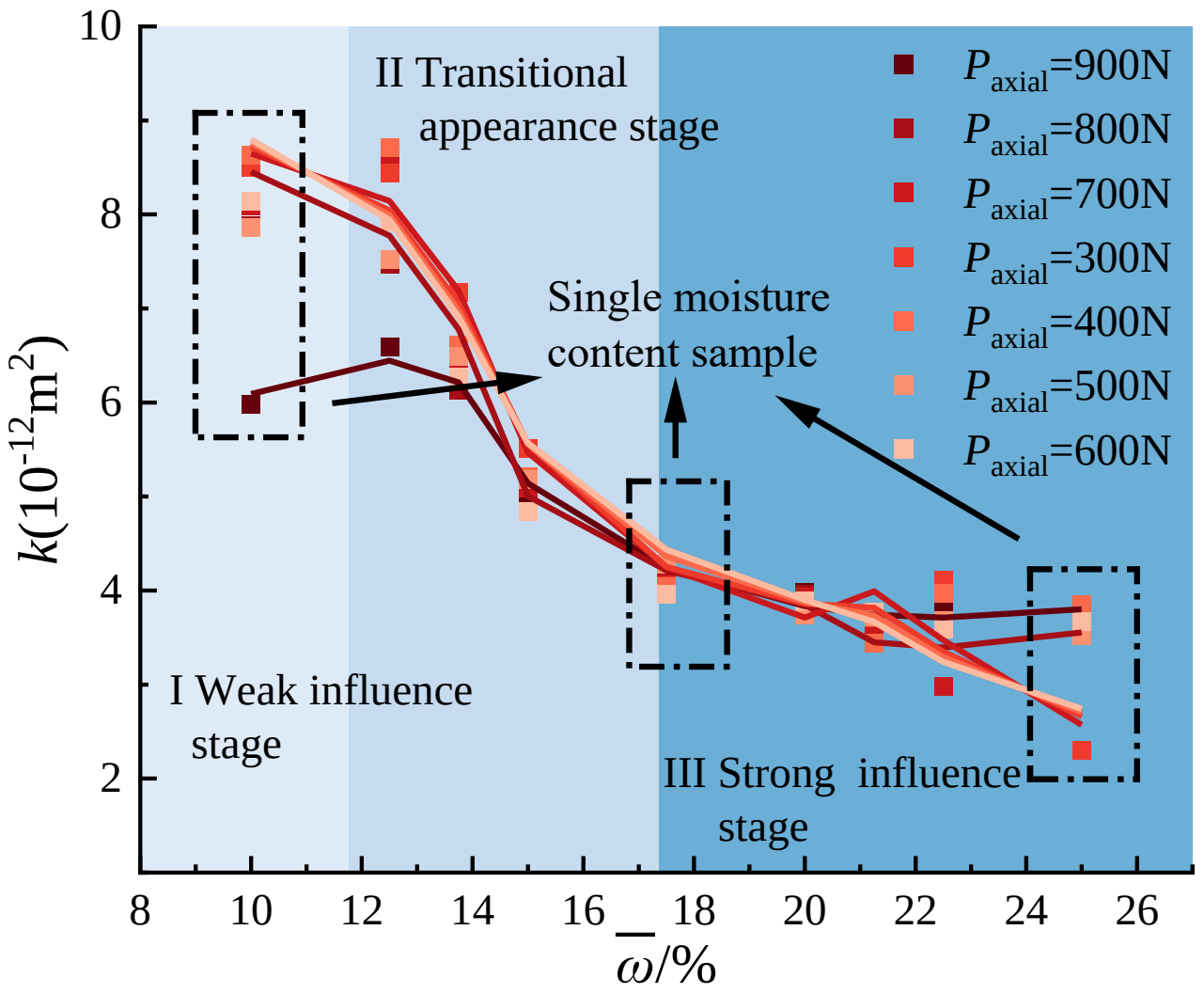

Fig.7 Variation curve of permeability with water content under different coaxial pressure($P_2$=0.075MPa)

It is found in Fig.7 that the permeability decreases gradually with the increase of (equivalent) water content under different axial pressures. With the change of water content, the influence of axial pressure on permeability shows positive and negative correlation modes. When the water content is low, the axial pressure is negatively correlated with the permeability. When the water content is high, it shows a positive correlation. This may be because at low water content, due to the increase of axial pressure, the porosity of the sample decreases, resulting in a decrease in permeability. At this time, water only fills part of the pores in the sample. At high water content, the polarization effect and axial compression effect act on the sample at the same time, but the water is more involved in the polarization effect, and the permeability increase caused by it is greater than the permeability decrease caused by axial compression.

With the increase of water content, the permeability affected by axial compression shows three stages : weak influence of polarization effect, transition of polarization effect and strong influence of polarization effect. When $\omega \le 10$ %, the polarization effect is weak, and the axial pressure has a significant inhibitory effect on the permeability, showing that the higher the axial pressure, the lower the permeability. When the polarization effect transition stage of 10 % < $\omega$ < 17.5 %, with the increase of water content, the polarization effect gradually appears, and the permeability of the sample under high axial pressure increases, showing that the negative correlation between axial pressure and permeability decreases. When $\omega \ge 17.5$ %, there is sufficient water in the sample to participate in the polarization process, and the polarization effect is fully apparent. The increase of permeability caused by the polarization effect completely compensates for and exceeds the decrease of permeability caused by axial compression. With the increase of axial compression, the permeability shows an increasing trend.

## 5 Discussion

### *5.1 Orientation of coal penetration*

According to equation (3), the permeability of positive and reverse sequence tests is calculated, as shown in figure 8. It is found that there are significant differences in the permeability calculated by the two types of permeability tests. Compared with the calculation results of the permeability in Reference [26], there are also significant differences in the permeability of the positive and reverse order tests, which confirms that the differences in this test are not accidental. It is speculated that there is a directional problem of permeability in the positive and reverse order permeability tests.

(1) Influence of osmotic pressure on permeability directivity

When $P_{axial}$ =300N, it is found in Fig.8 that, with the increase of equivalent water content and osmotic pressure, the difference between positive and reverse permeability

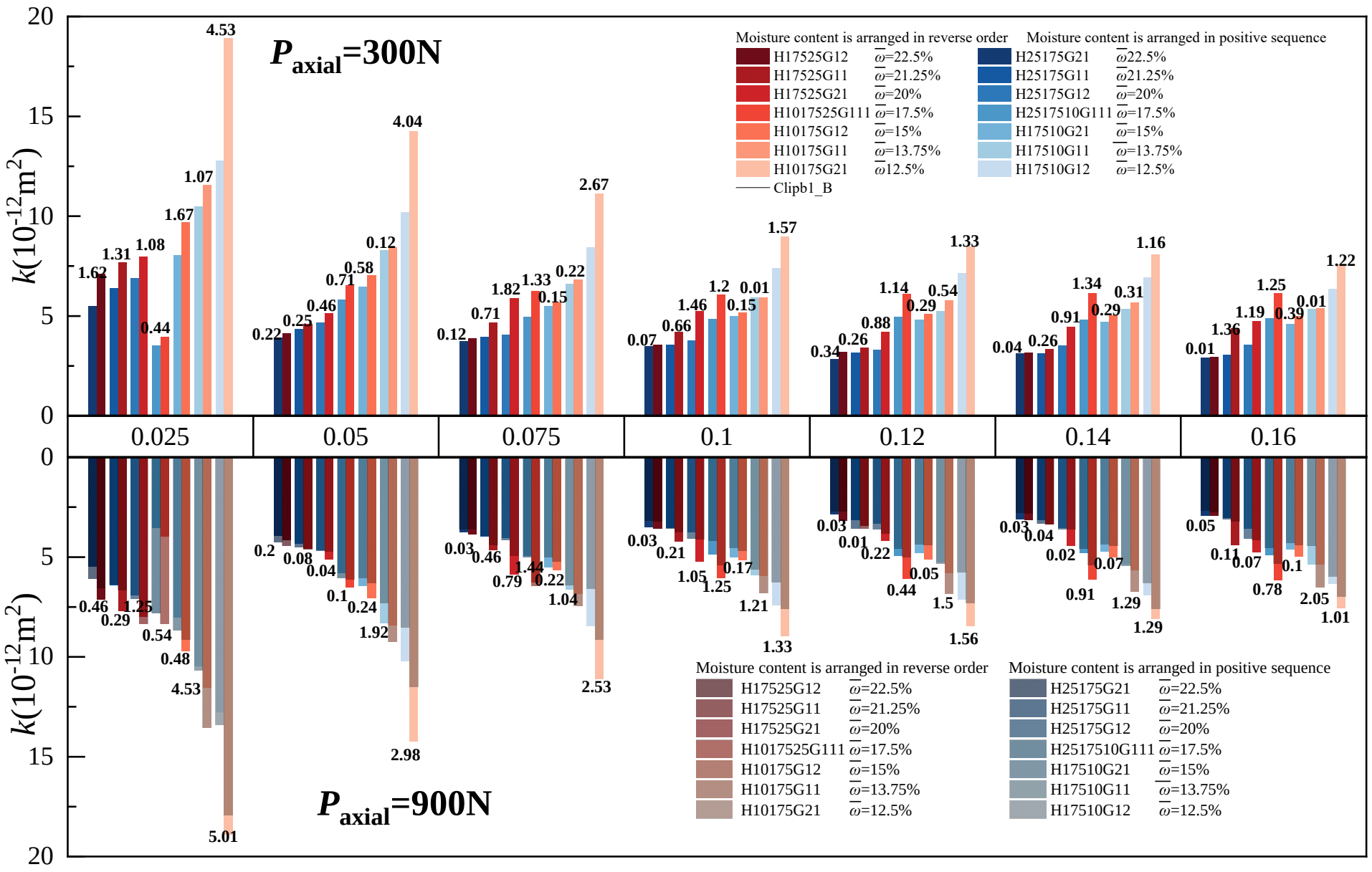

(a)

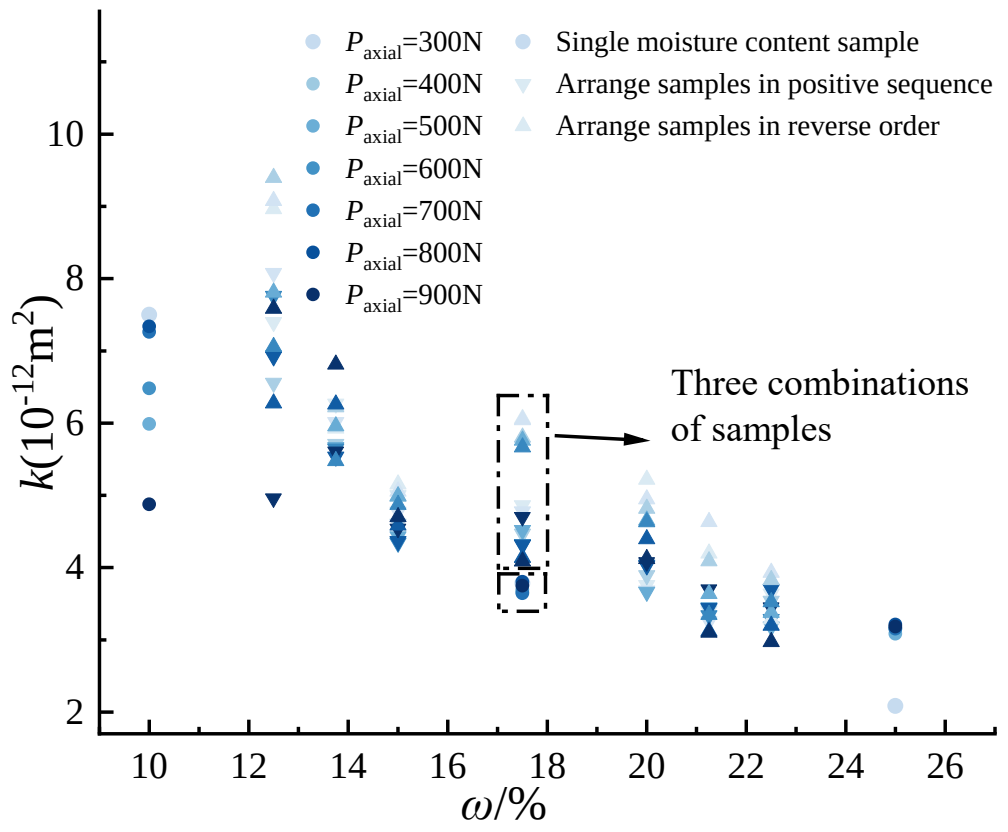

(b)ω-k (P2=0.1MP)

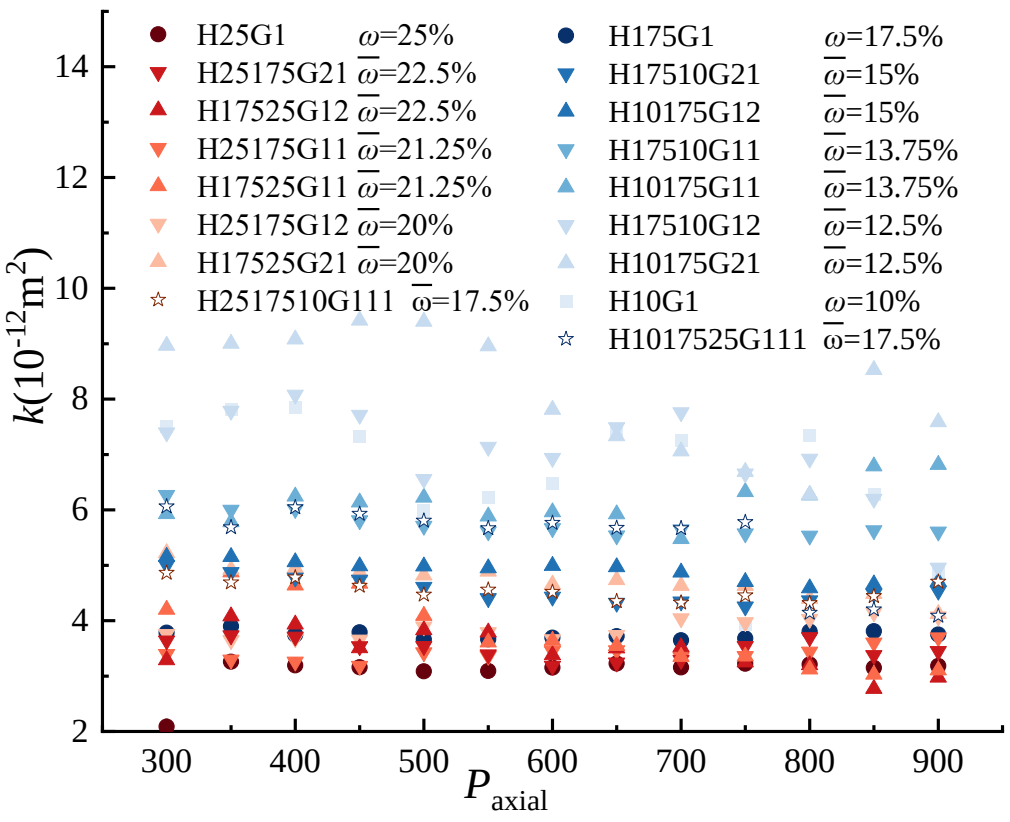

(c)$P_{axial}$-$k$（$P_2$=0.1MPa）

Fig.8 Relationship between osmotic pressure, moisture content, axial pressure and permeability of samples

gradually decreases. Comparing the data of $P_{axial}$ =300 N and $P_{axial}$ =900 N, it is found that with the increase of axial pressure, the difference between positive and reverse permeability decreases gradually. The directivity of permeability is affected by equivalent water content, osmotic pressure and axial pressure. The influence of water content is the first, the influence of osmotic pressure is the second, and the influence of axial pressure is less.

In the test of positive and reverse sequence, different permeability and directivity are shown. This may be due to the polarization effect of the sample in the process of infiltration, which leads to the existence of a lean water path in the sample. Therefore, in the positive and reverse osmosis processes, the gas will first pass through the lean water path and then pass through the high water content region and the low water content region respectively, resulting in the difference in positive and reverse permeability. When $P_{axial}$ =300 N and $P_{axial}$ =900 N, the difference between the positive and reverse sequence permeability of the sample gradually decreases, which means that the directivity is gradually weakened. This may be due to the multiple polarization effects occurring in the sample during the permeation process, and their superposition reduces the difference between the positive and reverse sequence permeability of the sample.

(2) Influence of sample moisture content on permeability directivity

In order to further describe the influence of water content on permeability directivity, the relationship between positive and reverse permeability and water content of the sample was drawn, as shown in Fig. 8(b).

Observing Fig.8(b), it is found that with the increase of (equivalent) water content, the permeability presents a nonlinear decrease and shows consistency under different axial pressures. Among them, the permeability of the three composite specimens is the lowest when the axial pressure is 300 N, which may be due to the existence of two interfaces with different upper and lower water content of the three composite specimens, and the polarization effect is not obvious when the three composite specimens have not undergone multiple permeation processes. At this time, the interface has a high obstruction effect on the seepage, so the permeability is the lowest.

Studies in literature [27] show that in the process of seepage, changes in the pore structure between the upper and lower parts of the double-pore composite coal will lead to flow resistance. In the samples of the three-pore composite, pore structure changes twice, resulting in increased flow resistance and low permeability. However, under other axial pressure conditions, the polarization effect has been produced due to the multistage osmotic pressure, so the reverse permeability is always higher than the positive permeability than the single water content.

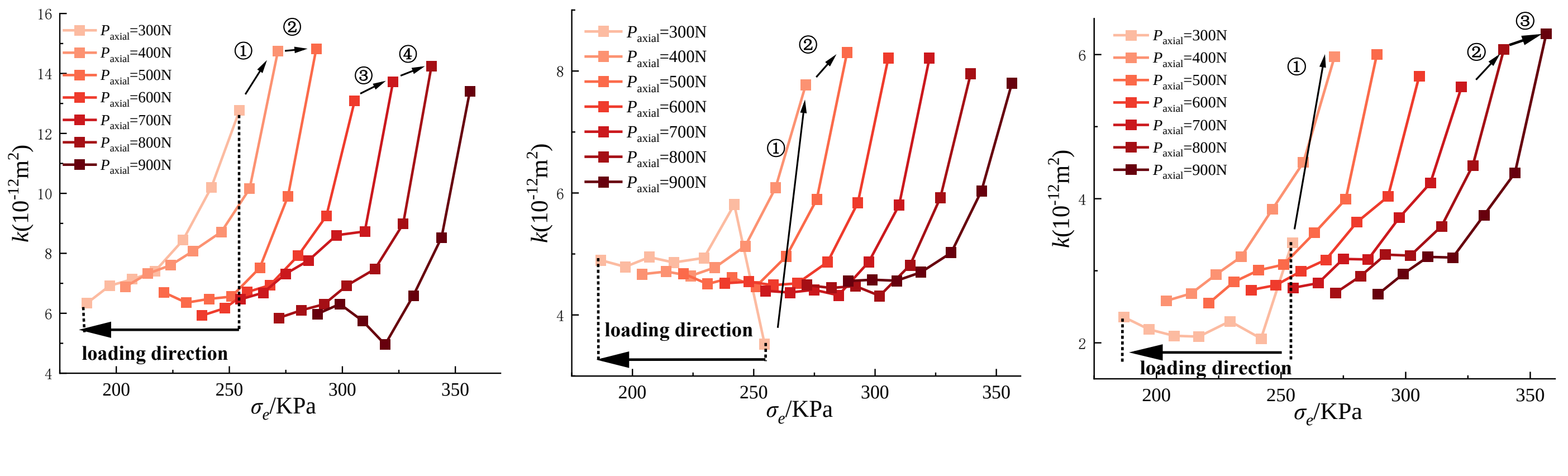

(a) H17510G12 (b) H2517510G111 (c)H25G1

Fig.9 Effective stress-permeability curve under multiple osmotic pressures

*5.2 Polarization effect in the process of penetration*

In order to represent the variation of the permeability of the sample during loading, the parameter of effective stress was used to characterize the loading of the sample, and the variation curve of effective stress-permeability of the three types of samples was drawn, as shown in Fig.9.

In Fig.9, it is found that the polarization effect will occur during the loading process of the three types of samples. In the same sample, with the loading of axial pressure and osmotic pressure, the polarization effect may occur many times. The higher the equivalent water content, the more obvious the increase of permeability caused by the polarization effect of the sample.

This may be due to the low water content of the sample after a single osmotic pressure loading has produced poor water passage; However, due to the high proportion of water in the pores of the samples with high water content, there is still a large amount of occupying water after one loading, and a stable water-poor path can be formed after multiple osmotic pressure loading. As shown in Fig.9(b), unsteady seepage occurs in the samples of the three combinations when the osmotic pressure is low. This may be due to the existence of two interfaces with different upper water content in the three composite samples, and the polarization effect is not obvious when the interface has not undergone multiple permeation processes. At this time, the interface has a high obstruction effect on seepage, so unstable seepage occurs and the permeability is low.

Under ideal conditions, with the increase of axial pressure, the internal porosity of the sample gradually decreased, the gas seepage channel decreased, and the permeability showed a downward trend. However, in Fig.9, axial pressure increases and permeability increases. It is speculated that the polarization effect may occur in the coal after applying multistage osmotic pressure. In the process of polarization effect, the water in the original water-containing coal sample occupies most of the gas seepage channels, forming placeholder water. Under the action of osmotic pressure, the gas drives away the occupying water and generates a seepage path conducive to gas passage. The water in the coal body changes from a disordered state to a directional distribution, forming two states of incomplete polarization and complete polarization, as shown in Fig. 10(a).

*5.3 Interaction between polarization effect and permeability directivity*

According to the polarization effect shown in Fig.10(a) and the discussion in Section 5, both directivity and polarization effects exist in the penetration test of the composite sample, and the process is shown in Fig.10(b).

As shown in Fig.10(b), the polarization effect generated during the positive and reverse permeability tests will cause the original water in the coal sample to move and change the seepage resistance. At this time, two states of incomplete polarization and complete polarization will be formed. In the case of incomplete polarization, the generated water-poor path has not penetrated the sample, and the gas still needs to pass through the water-poor path and flow through the water-containing part of the sample during the permeation process, showing that the permeability between the positive and reverse sequences of the sample is different.

At the same time, the polarization effect has the effect of weakening the permeability directivity of the sample. After several polarization processes, the permeability directivity of the sample completely disappeared, and the permeability between the positive and reverse sequences of the sample converged to form a completely polarized state.

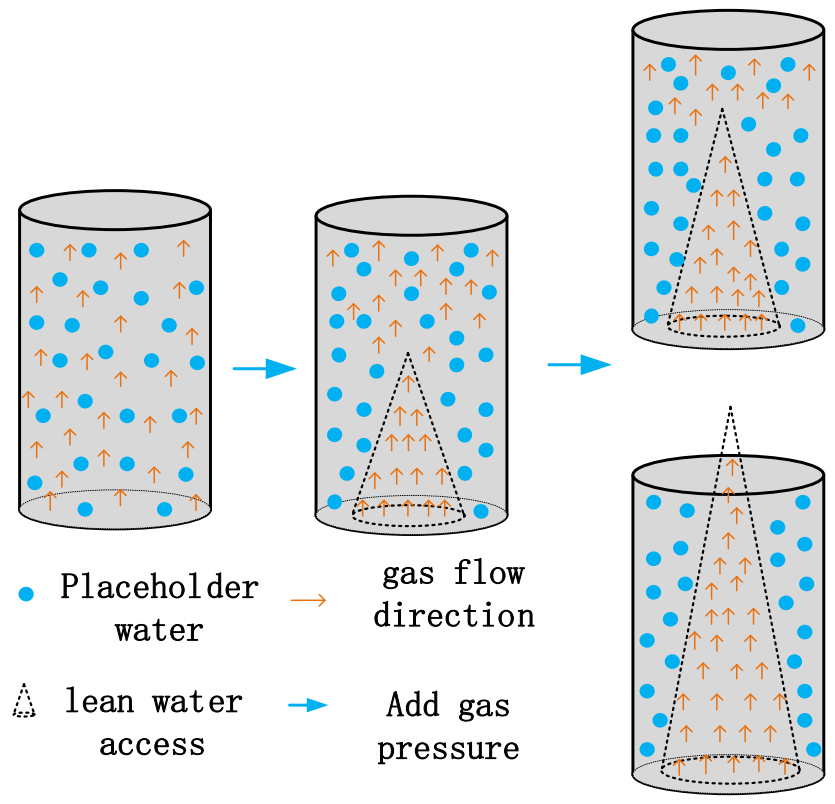

(a) Polarization effect process

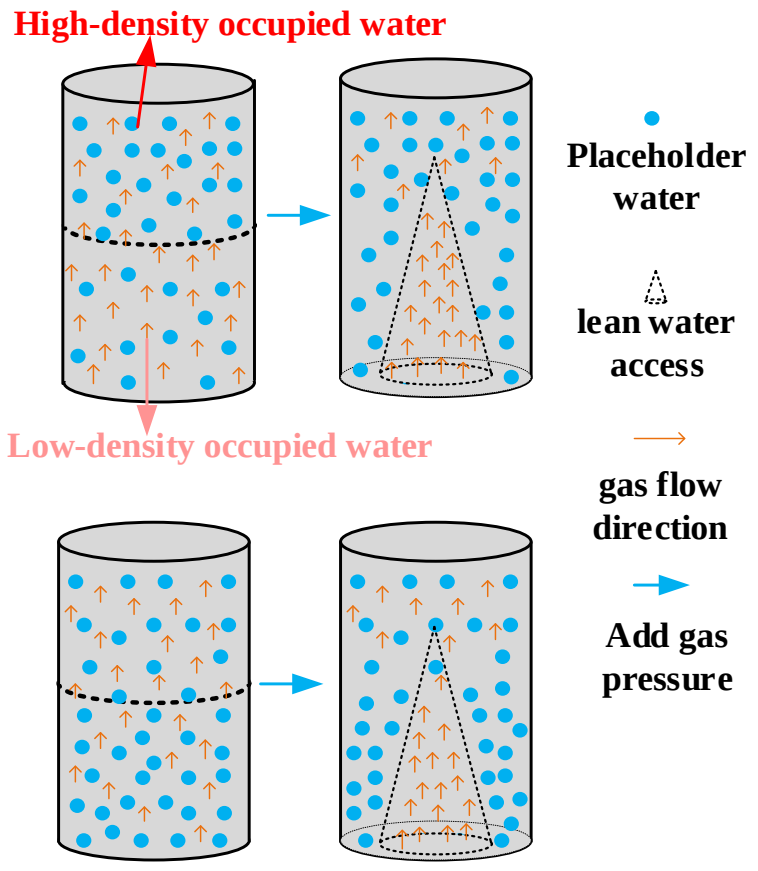

(b) Polarization effect and directivity

Fig.10 Polarization effect

## 6 Conclusions

In this paper, an improved triaxial osmometer was used to test the permeability characteristics of coal samples with different water content. The relationship between porosity, permeability, pressure gradient and seepage velocity of different samples was analyzed. The influence degree of water content on permeability, orientation and polarization of permeability were discussed, and the following conclusions were reached:

1) The relationship between permeability and porosity presents two trends of exponential and logarithmic type. These two curves divide the φ-k plane into three regions of super-exponential influence I, exponential influence II and logarithmic influence III, which respectively represent the trend of super-exponential, exponential and logarithmic increase of permeability with porosity.
2) As the absolute value of the pressure gradient increases, the permeability of the sample presents an exponential decay trend, and it shows two stages of slow and rapid decline. The higher the water content, the lower the initial permeability, the lower the permeance-pressure gradient characteristic curve as a whole moves down, the slow decline area moves to the right, and the absolute value of the pressure gradient in the region where the sample permeability enters a slow decrease decreases.
3) The seepage velocity of standard samples with different water content is negatively correlated with the pressure gradient. At all levels of osmotic pressure, the seepage velocity decreases with the increase of water content, but the decreasing trend is gradually weakened.
4) The directivity of permeability is affected by the equivalent water content, osmotic pressure and axial pressure, and the permeability of reverse sequence is generally higher than that of positive sequence. With the increase of equivalent water content, osmotic pressure and axial pressure, the difference between positive and reverse permeability decreases gradually, and the directivity weakens gradually. The influence of moisture content on directivity is the first, the influence of osmotic pressure is the second, and the influence of axial pressure is less.
5) There are three modes of influence on the polarization effect: weak influence, transition influence and strong influence, and there are two states: incomplete polarization and complete polarization. The polarization effect can weaken the permeability directivity. When the coal samples of the water-containing composite form a completely polarized state, the permeability directivity disappears.